# Volumetric optical scattering neural networks

Xuhao Luo[1, 2], Qiang Song[1, 3, *], Weiwei Cai[4], Lei Chen[5], Enbo Yang[4], Hao Wang[5], Zhipei Sun[6], Yueqiang Hu[1], Joel K. W. Yang[5], Cheng-Wei Qiu[2, 7, *] and Huigao Duan[1, 3, *]

[1]College of Mechanical and Vehicle Engineering, Hunan University, Changsha 410082, China.

[2]Department of Electrical and Computer Engineering, National University of Singapore, Singapore 117583, Singapore.

[3]Greater Bay Area Institute for Innovation, Hunan University, Guangzhou 511300, China.

[4]Key Laboratory for Power Machinery and Engineering of Ministry of Education, School of Mechanical Engineering, Shanghai Jiao Tong University, Shanghai 200240, China.

[5]Engineering Product Development, Singapore University of Technology and Design, Singapore 487372, Singapore.

[6]QTF Centre of Excellence, Department of Electronics and Nanoengineering, Aalto University, Espoo FI-00076 Aalto, Finland.

[7]National University of Singapore Suzhou Research Institute, Suzhou 215123, China.

*Correspondence to nanosong@hnu-gba.com (Q.S.); chengwei.qiu@nus.edu.sg (C.-W.Q.); duanhg@hnu.edu.cn (H.D.)

Optical neural networks rely on planar photonic circuits or spatially separated diffractive planes, limiting volumetric integration and requiring precise alignment. Here we propose a volumetric optical scattering neural network (OSNN) comprising densely packed refractive-index neurons with weak contrast $\Delta n = 0.006$. Direction-dependent first-order Born scattering provides local coupling, while ordered propagation accumulates these weak updates into global task-specific transformations. Fabrication-aware training and two-photon nanolithography produced a sub-$4\times10^{-4}$ $mm^3$ device containing 376,320 neurons at $1.0\times10^{9}/mm^3$ in a continuously written, intrinsically registered volume. Experimentally, a classifier performed class-selective energy routing with 95.8% blind-test accuracy on MNIST. The calculated passive optical transit time is below 1 ps. A separately optimized imager provided fourfold full-field spatial-sampling compression, resolved 1-μm features and enabled cross-domain transfer. An OSNN co-designed for multiple tasks encoded single-cell fluorescence-derived images into measured output maps, which were pooled into a 16-dimensional representation supporting organelle classification, morphology reconstruction, doublet detection and morphology regression, with all reported metrics exceeding 0.92. Volumetric weak scattering thus establishes an ultracompact, intrinsically registered optical computing architecture, opening a new route to scalable machine vision and biomedical analysis.

## Introduction

Artificial-intelligence workloads increasingly strain the energy and latency limits of conventional electronics[1–3]. Transistor scaling is constrained by heat, interconnect capacitance, and device physics[2]. Meanwhile, von Neumann data movement increasingly dominates energy use and latency[3]. These constraints motivate architectures that process information through physical propagation. During propagation, optical neural hardware can perform high-bandwidth, low-latency, parallel transformations[4,5]. Most implementations adopt one of two physical architectures. Photonic circuits perform linear transformations using coherent interference or wavelength-multiplexed processing[6–9]. By contrast, diffractive networks use patterned planes coupled by free-space propagation[10–12]. Photonic systems have demonstrated classification[6–12], imaging and feature extraction[10–12], spiking neurosynaptic processing[13], generative modeling[14], and large-scale optical interconnection[15]. However, their computation remains based on interference among photonic circuit elements or diffraction between separated planes.

Physical architecture determines network connectivity, volumetric density and registration requirements[4,5]. In interferometric photonic circuits, trainable elements and routing remain predominantly two-dimensional[5]. Diffractive networks propagate axially, but their trainable degrees of freedom are confined to separated surfaces[4,12]. Broad coupling commonly requires gaps of about $10^2$ wavelengths. These gaps leave much of the propagation volume without trainable degrees of freedom, while separately fabricated planes impose assembly and registration constraints[10–12]. Representative metasurface-based, nanoprinted, and vacancy-array spatial diffractive networks[16–23] achieve densities of $10^7$–$10^8$/mm$^3$. Fabrication resolution affects this ceiling, but separation between functional planes creates a more fundamental architectural bottleneck[12]. A continuously occupied trainable volume would instead co-distribute transformations and optical interconnections throughout a monolithic medium, increasing volumetric density while providing intrinsic registration. Yet neural computing has not fully exploited the intrinsic three-dimensional evolution of light as the computational substrate.

Here we introduce and experimentally realize the OSNN, which distributes weak refractive-index perturbations with $\Delta n = 0.006$ throughout a monolithic three-dimensional volume. Each perturbation acts as a neuron whose direction-dependent first-order Born scattering couples mainly to a local neighborhood. Ordered propagation accumulates these weak local updates into global, task-specific transformations. We combine a differentiable weak-scattering model with fabrication-aware training and two-photon nanolithography. Using this approach, 376,320 trainable voxels are packed at $1.0\times10^9$/mm$^3$ in a monolithic volume that requires no inter-plane alignment. In simulation, the locally connected OSNN achieved 96.52% MNIST[24] classification accuracy over a 180-μm optical path, compared with 90.26% over 816 μm for the optimized fully connected diffractive neural network (DNN) benchmark. The calculated passive optical propagation time is as short as 0.82 ps. Experimentally, the fabricated classifier achieved 95.8% blind-test accuracy. A separately optimized imager provided fourfold full-field spatial-sampling compression and resolved 1-μm features, with cross-domain transfer from Fashion-MNIST[25] (FSIM 0.93) to VesselMNIST3D[26] (FSIM 0.91). An OSNN co-designed for multiple tasks produced a 16-dimensional single-cell representation that supported organelle classification (0.96 accuracy), morphology reconstruction (SoftDice 0.949), doublet detection (TPR 0.965 at 10% FPR), and morphology regression (coefficient of determination 0.926) through task-specific electronic decoding. These results establish volumetric weak scattering as a high-density, intrinsically registered architecture in which repeated local interactions build global optical functions over a short propagation distance.

## Physical basis and local connectivity

An OSNN densely distributes weak refractive-index perturbations throughout a transparent three-dimensional medium. As illustrated in Fig. 1a, each perturbation functions as an optical neuron whose response depends on its refractive index, geometry, and incident field. We model each neuron as a cubic voxel with edge length $l_v$, although the formalism is geometry-independent and extends directly to other scatterer shapes, including spheres, prisms, and cones. The small refractive-index contrast justifies a first-order Born treatment of each local scattering update[27], with local scattering potential $V_j = k_0^2\left(n_j^2 - n_b^2\right)$. Here, $n_j$ and $n_b$ denote the neuron and background indices, respectively, and $k_0 = 2\pi/\lambda_0$. A single voxel weakly perturbs the field, whereas successive perturbations redistribute optical energy over a short path. The accumulation of these weak local updates underpins the OSNN. The architecture is not inherently restricted to weak scattering. Extending it to intermediate or strong scattering requires a full multiple-scattering treatment that captures higher-order local interactions and backward or recurrent coupling in the forward and gradient calculations.

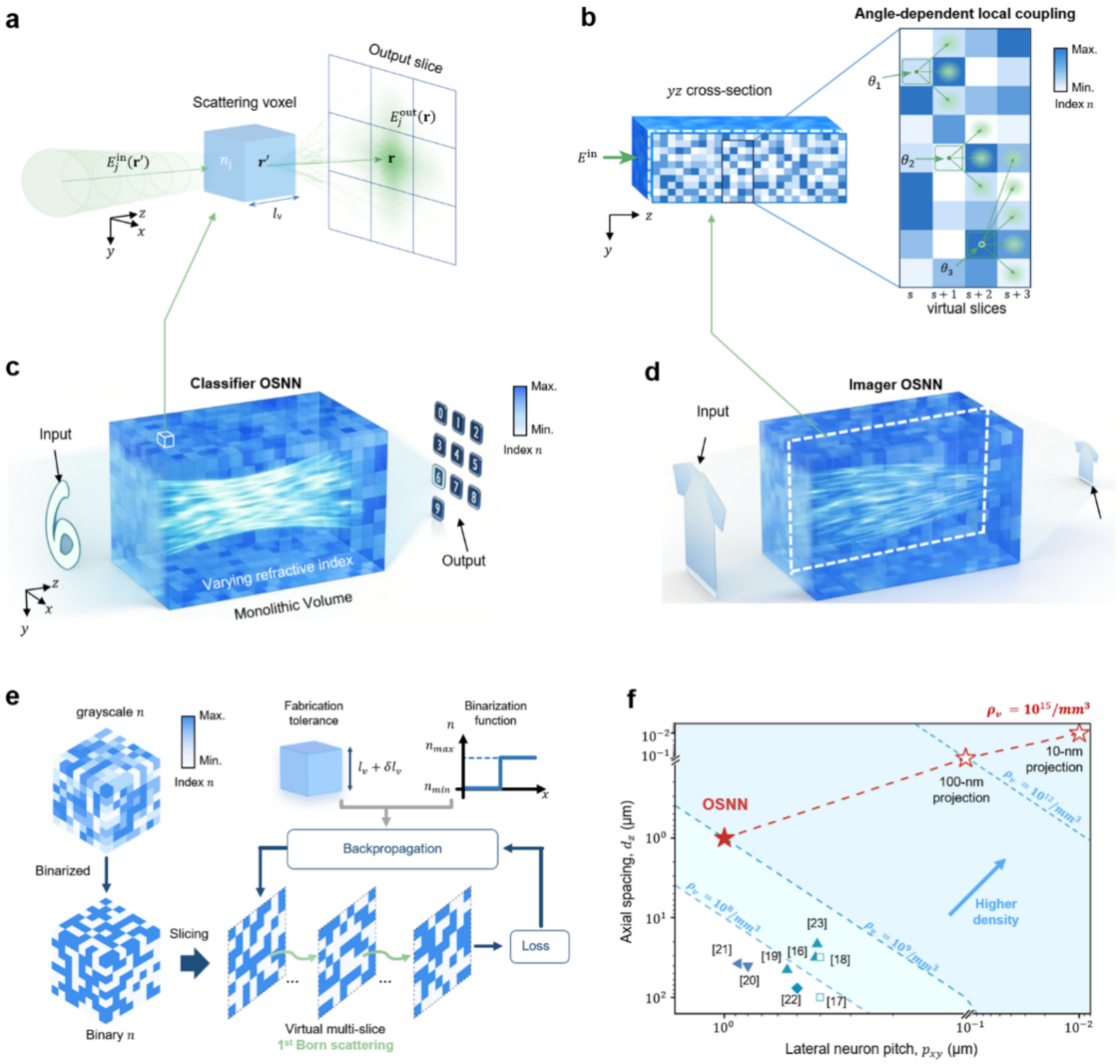

***Fig. 1 | Principle and architecture of a volumetric optical scattering neural network. a,*** *Direction-dependent first-order scattering from a refractive-index perturbation under the first Born approximation.* ***b,*** *Virtual-slice representation of local, angle-dependent coupling, in which successive propagation and weak-scattering updates accumulate across the volume.* ***c, d,*** *Task-specific OSNNs for class-selective energy routing (c) and compressed imaging (d).* ***e,*** *Fabrication-aware inverse design using slice-wise backpropagation, binary index projection, and a global isotropic dimensional bias $\delta l_v \in [-10, 10]$ nm.* ***f,*** *Map of lateral pitch and axial spacing for representative experimental transverse-spatial optical neural networks[16–23]. Axes are reversed so*

*smaller pitches appear toward the upper right. Density contours show* $\rho_v = 10^9/(p_x p_y d_z)$ *for dimensions in micrometers. The filled star marks the demonstrated OSNN (1-μm isotropic pitch;* $1.0\times10^9/\text{mm}^3$*). The open star schematically marks a theoretical isotropic 10-nm-grid projection* ($1.0\times10^{15}/\text{mm}^3$*), based on the reported fabrication feature scale (see Supplementary Note 3). Each independently optimized modulating or scattering element is counted as one neuron.*

For a first-order Born treatment of the local voxel response, the field produced by neuron $j$ at observation position $\mathbf{r}$ is

$$E_j^{\text{out}}(\mathbf{r}) = E_j^{\text{in}}(\mathbf{r}) + \int_{\Omega_j} G_b(\mathbf{r},\mathbf{r}')V_j E_j^{\text{in}}(\mathbf{r}')\,\mathrm{d}^3\mathbf{r}', \tag{1}$$

where $E_j^{\text{in}}(\mathbf{r}')$ and $E_j^{\text{out}}(\mathbf{r})$ denote the incident field within neuron $j$ and the total field at $\mathbf{r}$, respectively. The neuron volume is $\Omega_j$, and $G_b(\mathbf{r},\mathbf{r}')$ is the Green function of the homogeneous background. Equation (1) therefore provides a differentiable single-neuron response.

The virtual-slice construction in Fig. 1b discretizes the monolithic volume along $z$ into slices of thickness $\Delta z = l_v$. The field transfer from slice $s$ to $s+1$ is

$$E_{s+1}(\mathbf{r}_\perp) = \mathcal{P}_{\Delta z}[E_s](\mathbf{r}_\perp) + \mathcal{G}_{\Delta z}[V_s E_s](\mathbf{r}_\perp), \tag{2}$$

in which $\mathbf{r}_\perp = (x, y)$, while $E_s(\mathbf{r}_\perp)$ and $E_{s+1}(\mathbf{r}_\perp)$ denote the fields on adjacent virtual slices. The operator $P_{\Delta z}$ propagates the incident field through the background over $\Delta z$, whereas $G_{\Delta z}$ adds the field generated by the slice potential $V_s(\mathbf{r}_\perp)$. Equation (2) combines background propagation with the current weak-scattering update. Iterating Eq. (2) yields the slice-resolved internal fields shown in Fig. 2c,d. The virtual slices are numerical sampling planes rather than fabricated layers and therefore require no inter-slice alignment.

The effective OSNN scattering footprint is local and direction dependent, while its excitation depends on the input field. Each neuron receives a coherent superposition from upstream voxels and couples most strongly to a nearby downstream region. Neither the Green function nor the scattered field has a strict finite range, but the computed finite-voxel response is concentrated within a local downstream footprint. Supplementary Note 1 and Supplementary Fig. S1 indicate an effective 3×3 to 5×5 neighborhood at normal incidence that shifts and broadens under oblique illumination. Under a 5×5-neighborhood estimate, a path can advance by up to two lateral voxels per slice, giving a geometric reach of approximately 56 voxels after 28 updates. Supplementary Fig. S2 provides a task-specific numerical criterion for sufficient propagation depth to realize the learned global transformation. This locality resembles a receptive field, but the couplings arise from scattering physics and do not share a digital convolution kernel.

We optimize the three-dimensional refractive-index distribution by backpropagating the task loss through the optical model. The optimization is constrained by the ordered weak-scattering model, local coupling, the accessible refractive-index range, and fabrication tolerance. The reported devices use two index states for reproducible fabrication, although dose control can produce grayscale distributions. We count each independently optimized scattering voxel as one optical neuron. The density comparison in Fig. 1f uses $\rho_v = 10^9 \times \left(p_x p_y d_z\right)^{-1}$, with density in $mm^{-3}$ and lateral pitches and axial spacing expressed in micrometers. This operational definition enables reproducible density comparisons without implying identical transfer functions or computational capabilities.

Fig. 1c shows the classifier OSNN, which routes optical energy to ten class-specific detector regions. Fig. 1d depicts the imager OSNN, which remaps spatial information to a downsampled intensity output. Both devices use identical 56×56×120 geometries. In each system, an encoded complex field traverses the trained volume and, when used, a free-space output region before detection. Propagation is linear in the input field for a fixed material distribution. Square-law readout, $I_{\mathrm{out}} \propto |E_{\mathrm{out}}|^2$, provides the nonlinear measurement used for the loss and output decision.

The representative classifier and imager each comprise 56×56×120 cubic voxels with an edge length of $l_v = 1$ μm. The resulting 56×56×120 $\mu m^3$ volume contains 376,320 trainable neurons and 120 virtual slices. Increasing the axial extent provides more scattering updates and improves performance in the evaluated depth studies. We optimize latent refractive-index variables with Adam by backpropagating through Eq. (2). We observed no vanishing or exploding gradients during the 120-slice optimization. This stability is consistent with the residual-like form of Eq. (2), in which the background-propagated field provides a direct propagation path between slices, while each weak-scattering correction is small, limiting multiplicative gradient contraction or amplification.

The fabrication-aware workflow in Fig. 1e incorporates the material states and dimensional tolerances of dose-controlled two-photon lithography. At 532 nm, the unexposed IP-Dip2 matrix and exposure-modified regions have refractive indices of 1.547 and 1.553, corresponding to $\Delta n = 0.006$ between the two material states. Continuous latent variables are projected onto these states using a *binary tanh* projection. A straight-through estimator preserves gradient-based updates. At each optimization iteration, robustness training samples one global isotropic dimensional offset $\delta l_v \in [-10, 10]$ nm from this closed interval and applies it to the voxel edge length and corresponding lattice pitch in all three dimensions. Thus, every iteration exposes the optimizer to an arbitrary global isotropic dimensional bias within the full tolerance interval.

Because trainable scatterers and their interconnections occupy the same monolithic volume, local connectivity directly increases neuron density. Isotropic 1-μm pitches place 376,320 neurons in $3.7632\times10^{-4}$ mm$^3$, equivalent to $1.0\times10^9$/mm$^3$. Nanoprinted diffractive networks can use submicrometer lateral elements, but their axial separations remain tens of micrometers. Consequently, their reported volumetric densities remain in the $10^7$–$10^8$/mm$^3$ regime. By matching the axial and voxel pitches, the OSNN overcomes this architectural bottleneck and raises volumetric neuron density into the $10^9$ regime. At matched submicrometer pitch, this geometry could approach densities nearly two orders of magnitude higher than the densest separated-plane systems. A feature size of approximately 9 nm has been reported in three-dimensional optical beam lithography[28]. On the basis of this demonstrated feature scale, an isotropic 10-nm grid would provide a theoretical geometric design-site density of $1.0\times10^{15}$/mm$^3$, seven orders of magnitude above the $10^8$/mm$^3$ regime. This estimate assumes that the reported feature scale can be extended to a uniform three-dimensional lattice. The geometry definition, design-site-density calculation and literature audit are detailed in Supplementary Note 3.

## Weak-to-strong field transformation

Fig. 2a displays the optimized binary index distribution and orthogonal sections of the classifier, whereas Fig. 2b shows those of the imager. Both devices have identical 56×56×120 geometries. The classifier contained finer, more dispersed features, whereas the imager exhibited stronger axial correlations and channel-like motifs. These task-specific patterns are consistent with energy routing and spatial remapping, respectively.

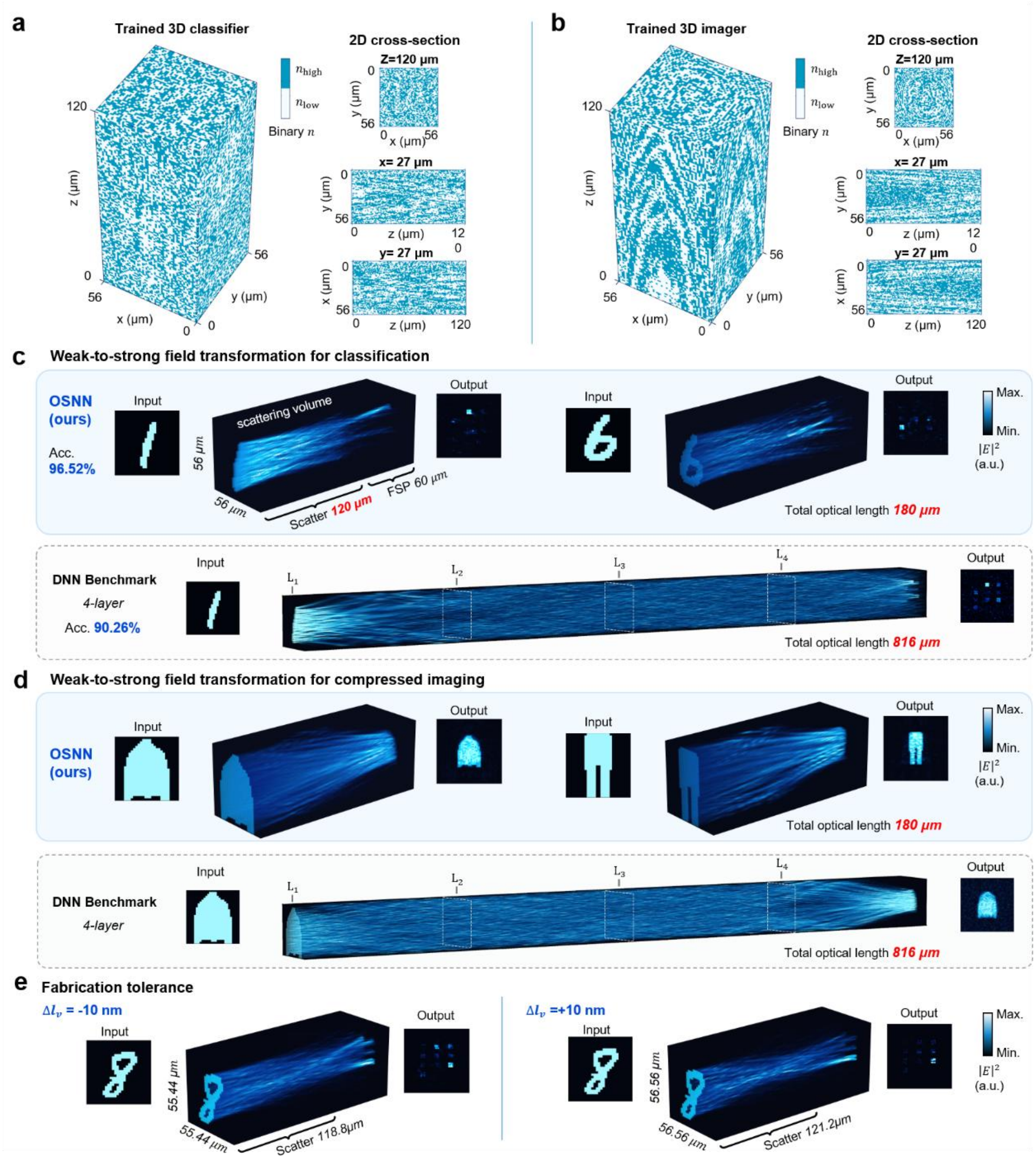

*Fig. 2* | ***Weak-to-strong field transformation in trained OSNNs. a, b,*** *Optimized binary index distributions and orthogonal sections of the classifier (a) and imager (b), with* $n = 1.547$ and $1.553$ *at a wavelength of 532 nm.* ***c, d,*** *Simulated normalized-intensity evolution for representative MNIST classification (c) and Fashion-MNIST compressed imaging (d). An optimized, fully connected DNN is included as a benchmark for task performance and optical-path length, not as an architecturally equivalent control. The OSNN contains a 120-μm scattering volume and a 60-*

*μm output-propagation region (180 μm total); the DNN spans 816 μm.* ***e,*** *Classifier outputs for digit '8' under uniform isotropic dimensional biases of −10 and +10 nm. FSP, free-space propagation.*

The MNIST classifier map in Fig. 2c and the Fashion-MNIST imager map in Fig. 2d visualize the normalized internal intensity. We obtained these maps by propagating the complex field slice by slice with Eq. (2). Each OSNN combines a 120-μm scattering volume with 60 μm of free-space propagation, giving a total optical path of 180 μm. The output-propagation region is configurable and is not intrinsic to the OSNN. Although each slice weakly perturbed the complex field, 120 successive updates strongly redirected and reshaped the optical energy. Classification inputs developed class-dependent flows toward target detector regions, whereas imaging inputs retained and remapped spatial structure. Thus, dense repetition converted weak scattering into a task-specific transformation that used the full propagation path. Supplementary Fig. S2 quantifies this mechanism in a long-range routing task. The mean squared-norm source-update ratio is ~0.245%, whereas the cumulative transformation reaches 0.913 at the volume exit and the 90%-energy footprint expands from ~7.4 to 45.2 μm. Depth and axial-order controls further show that global routing emerges only after sufficient propagation depth and is strongly suppressed by shuffling or reversing the trained volume.

We used an optimized, fully connected DNN to benchmark task performance and optical-path efficiency, not as an architecturally equivalent control. The locally connected OSNN achieved 96.52% simulated classification accuracy over 180 μm. The DNN benchmark achieved 90.26% over 816 μm. Thus, the OSNN gained 6.26 percentage points while shortening the optical path by 636 μm, or 78%. The benchmark reorganizes the field between separated planes, whereas the OSNN updates it at micrometer-scale intervals. Under the tested aperture, task, and optimization conditions, distributed local coupling used the available path more effectively than this specific benchmark.

During robustness training, a global isotropic dimensional bias $\delta l_v$ was sampled at every optimization iteration from the closed interval −10 to +10 nm and applied to the voxel size and corresponding lattice pitch in all three dimensions. As evidenced by the endpoint outputs in Fig. 2e, the maximum output for digit '8' remained within the target region at both offsets. Internal and fine-scale output patterns nevertheless changed. The correct target region was therefore preserved at both endpoints of the trained tolerance interval.

## Experimental class-selective routing

In the experimental optical system shown in Fig. 3a, a spatial light modulator encoded each digit onto a 532-nm beam. A 4f system imaged the encoded field onto the OSNN input, and an objective collected the output for recording on a CMOS sensor. We classified each image by integrating the power within ten predefined detector regions. Intensity detection implements $I \propto |E|^2$, which is included in the forward model.

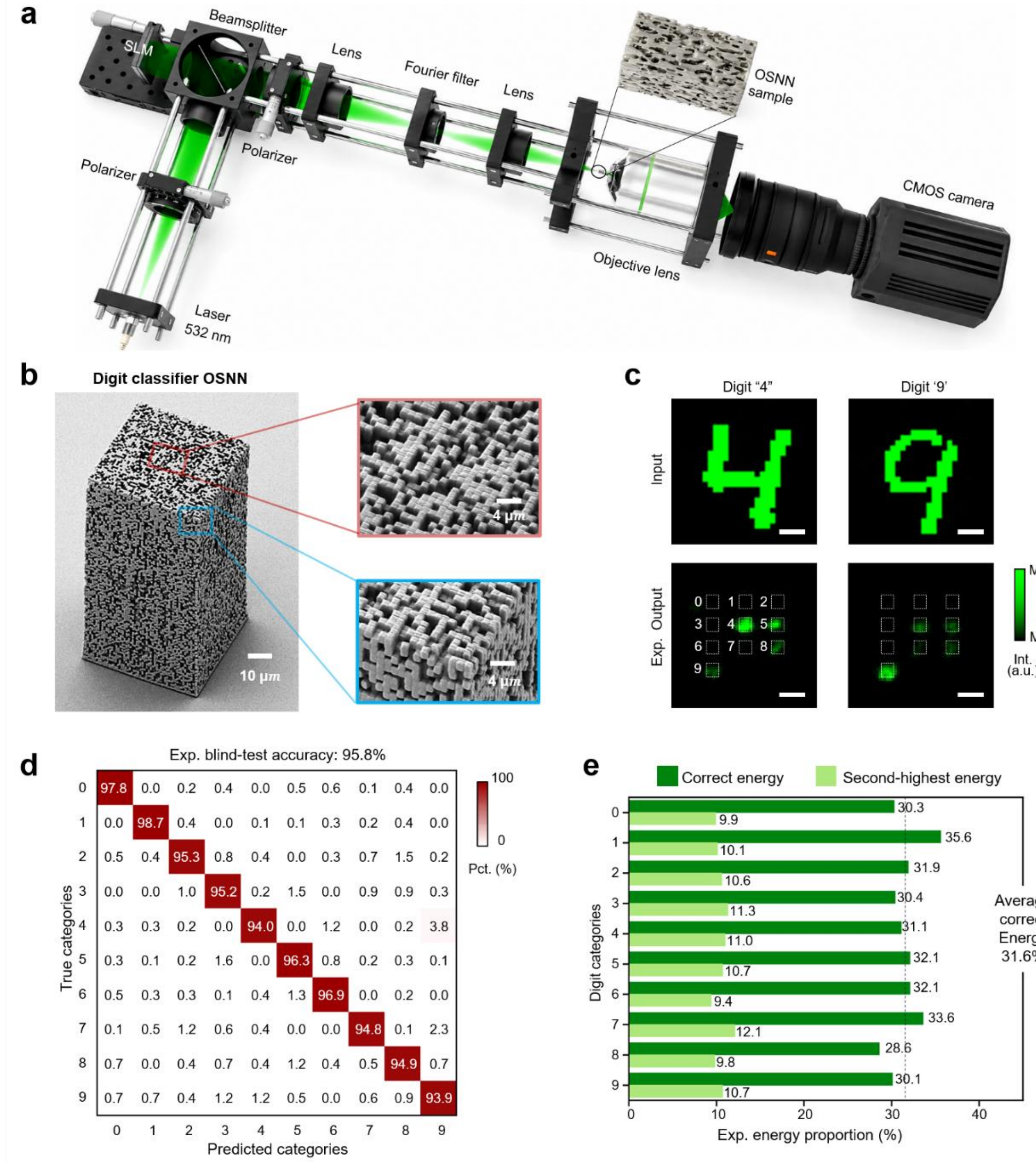

***Fig. 3 | Experimental class-selective energy routing in a monolithic OSNN. a,*** *Measurement system comprising 532-nm illumination, spatial-light-modulator encoding, a 4f system, and CMOS detection.* ***b,*** *SEM image and voxel morphology of the developed classifier. Optical measurements were acquired before development, while the device remained embedded in IP-Dip2.* ***c,*** *Measured outputs for digits '4' and '9'. Scale bar, 10 μm.* ***d,*** *Confusion matrix for 10,000 MNIST test images (95.8% blind-test accuracy).* ***e,*** *Correct-region and second-highest-region energies for each class.*

*The mean correct-region fraction is 31.6%.*

We printed the classifier in IP-Dip2 on a quartz substrate and measured it while it remained embedded in the surrounding matrix. We developed the device only afterward to expose its three-dimensional morphology for SEM. Fig. 3b shows the voxelized binary architecture, which closely matches the design in Fig. 2a. The classifier occupies $3.7632\times10^{-4}$ $mm^3$ and contains 376,320 co-written neurons and optical couplings, corresponding to $1.0\times10^9/mm^3$. This density requires neither separate planes nor inter-plane alignment. It therefore arises from the monolithic local-connectivity architecture rather than from lithographic pitch alone. Fabrication, undeveloped-state optical testing and post-test structural characterization are detailed in Supplementary Note 4.

Representative outputs for digits '4' and '9' in Fig. 3c directed the largest integrated energy to the target regions. The confusion matrix in Fig. 3d shows 95.8% blind-test accuracy across all 10,000 MNIST test images. For every class, the mean correct-region energy exceeded the mean second-highest-region energy. Fig. 3e summarizes these margins, with a mean normalized correct-region fraction of 31.6%.

## Compressed imaging and cross-domain test

In the compressed-imaging experiment, a separately optimized OSNN mapped each 56×56 input to a 28×28 output while preserving the full image extent. This mapping reduced the number of spatial samples from 3,136 to 784, providing fourfold full-field spatial-sampling compression. We fabricated the device using the same two-photon process as the classifier and measured it while embedded in IP-Dip2. The developed specimens illustrated in Fig. 4a were used only for SEM. One specimen omits one-eighth of the volume to expose the internal voxel pattern.

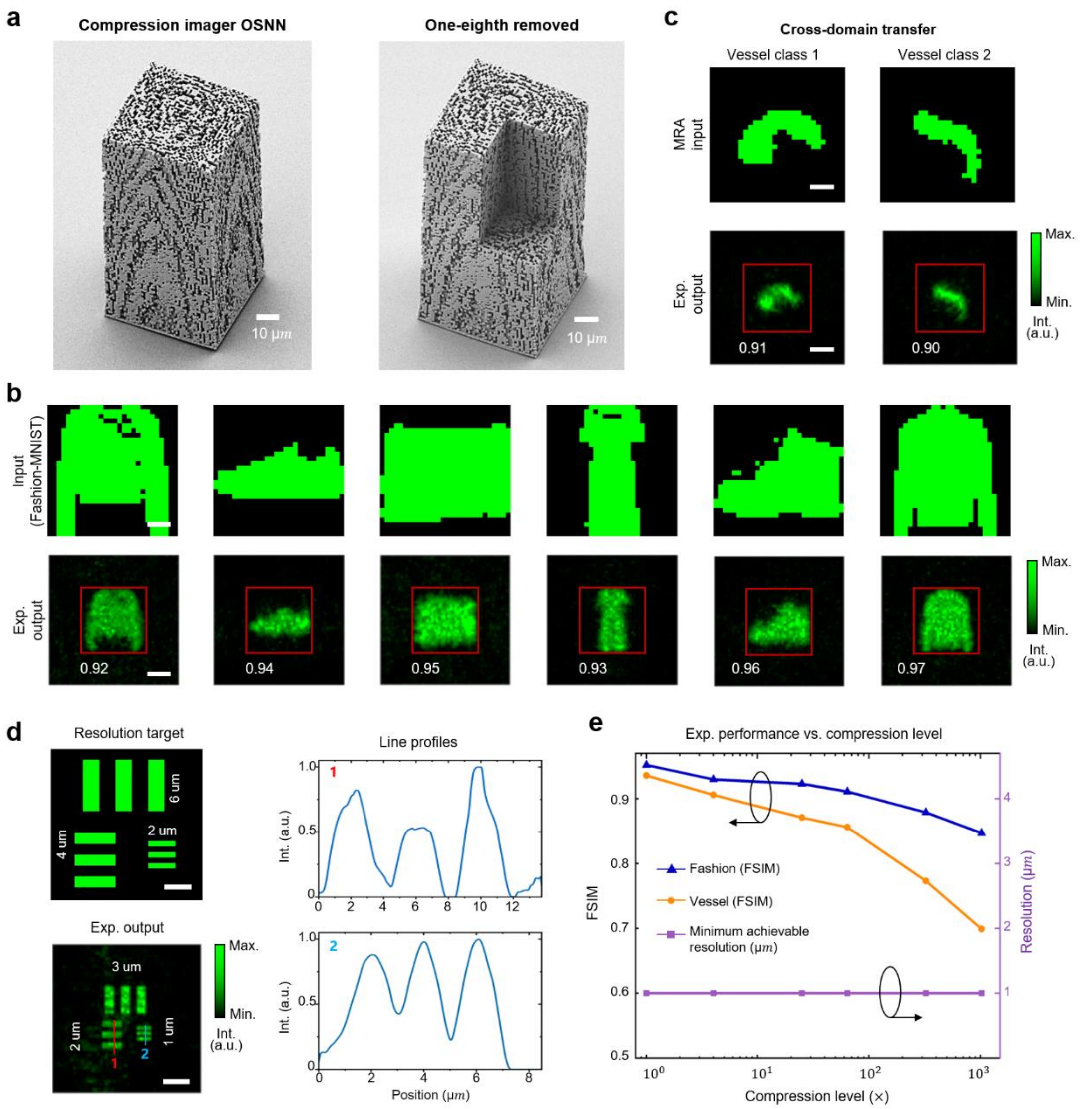

*Fig. 4 | **Experimental compressed imaging, cross-domain transfer, and resolution.** **a,** SEM images of developed imager structures. One specimen omits one-eighth of the volume to expose the interior; the measured device remained embedded in IP-Dip2. **b,** Fashion-MNIST reconstructions and measured feature similarity index (FSIM) values. **c,** VesselMNIST3D-derived reconstructions from*

*the same fixed OSNN without optical retraining.* ***d,*** *Resolution-chart output and line profiles. The 1-μm linewidth group satisfies the adopted line-separation criterion.* ***e,*** *Experimental mean FSIM and minimum resolved linewidth versus spatial-sampling compression (1–1024). Each point represents a separately optimized and fabricated OSNN with transverse dimensions selected for the target compression level; the sweep is not restricted to the 56×56 geometry used in b–d. Scale bars, 10 μm.*

Fig. 4b shows that, across 10,000 Fashion-MNIST[25] test images, the measured outputs preserved the dominant silhouettes and full spatial extent of each class. The mean feature similarity index (FSIM) was 0.93. We then tested the same fixed imager on 382 VesselMNIST3D-derived 56×56 projections from MedMNIST[26], without retraining, redesign, or refabrication. The reconstructions in Fig. 4c retained the dominant vessel contours. Mean FSIM decreased from 0.93 to 0.91, indicating cross-domain transfer of the learned physical mapping under the tested conditions.

We measured spatial resolution using a custom line chart. Adjacent responses were considered resolved when the intervening minimum satisfied $I_{\min} \leq 0.735(I_1 + I_2)/2$. The output and line profiles in Fig. 4d show that the 1-μm linewidth group met this adopted line-separation criterion, demonstrating resolvability of 1-μm features under the present input-sampling and collection conditions.

Fig. 4e summarizes the performance of separately optimized and fabricated imagers with compression factors from 1 to 1024. In this device-level scaling series, the transverse OSNN dimensions were varied to realize the target spatial-sampling ratios, rather than fixed at the 56×56 geometry of the representative 4× imager in Fig. 4b–d. Mean FSIM decreased with compression in both image domains because fewer samples represented the full field. Nevertheless, the 1-μm linewidth group remained resolved in every tested device. Thus, full-field similarity decreased with compression while the tested 1-μm features remained distinguishable. These measurements distinguish the two readout criteria without establishing that resolution is generally independent of compression.

## Single-cell fluorescence representation

In the single-cell experiment shown in Fig. 5a, we used an OSNN co-designed for multiple tasks as a shared optical encoder for hybrid single-cell fluorescence analysis. The fluorescence-derived images were encoded on the SLM under coherent 532-nm illumination. The shared refractive-index distribution was co-optimized across the four downstream objectives and then fixed for fabrication and experimental evaluation. The OSNN transformed each 56×56 input. Pooling the measured output over a fixed 4×4 grid then yielded a 16-dimensional vector. This pooling represents 3,136 measured spatial samples with 16 downstream channels, a 196-fold reduction in representation dimensionality. We passed the same vectors to four independently parameterized, task-specific electronic decoders for organelle classification, morphology reconstruction, doublet detection, and morphology regression. The fixed optical stage forms the representation, and spatial pooling reduces the measured output to 16 channels before task-specific nonlinear electronic decoding.

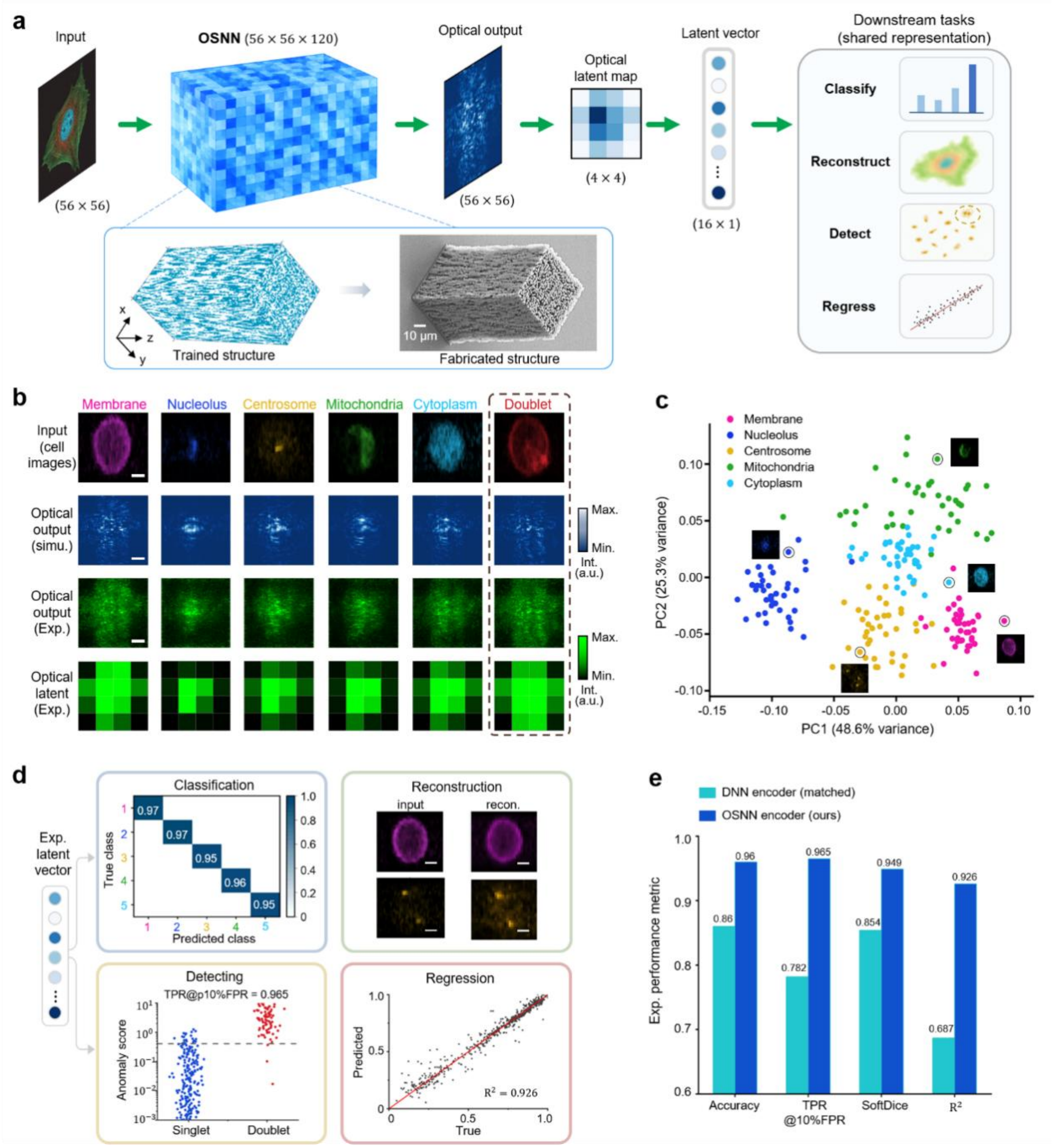

***Fig. 5 | A shared OSNN-encoded physical representation for hybrid single-cell fluorescence analysis. a,*** *An OSNN co-designed for multiple tasks maps each 56×56 image to a 16-dimensional representation through pooling over a fixed 4×4 grid, yielding 16 downstream channels from 3,136 measured spatial samples (196-fold dimensionality reduction). Task-specific electronic decoders perform organelle classification, morphology reconstruction, doublet detection, and morphology regression from the same OSNN-encoded physical representation. Insets show the optimized and fabricated structures.* ***b,*** *Inputs, simulated and measured outputs, and pooled latent maps for five*

*organelle classes and synthetic doublets.* ***c,*** *Principal-component analysis of experimental representations. PC1 and PC2 explain 48.6% and 25.3% of the variance, respectively.* ***d,*** *Cross-task performance, with 0.96 accuracy; a 0.965 true-positive rate (TPR) at a 10% false-positive rate (FPR); 0.949 SoftDice; and* $R^2 = 0.926$*.* ***e,*** *Comparison with an optical-path-length-matched DNN benchmark using identical output sampling, detector readout, and downstream decoder architectures. Scale bar, 10 μm.*

We used the EMBL-EBI BioStudies dataset S-BSST644[29], which comprises membrane, nucleolus, centrosome, mitochondria, and cytoplasm patterns. We used synthetic doublets as anomaly samples, motivated by prior optical detection of doublet-cell anomalies[30]. The representative measured 4×4 maps in Fig. 5b retained class- and doublet-dependent structure, despite compressing 3,136 measured spatial samples into 16 latent channels. In the principal-component projection in Fig. 5c, the first two components explained 73.9% of the variance and formed compact, class-dependent groups.

The four tasks probe discriminative, spatial, anomaly-sensitive, and continuous information in the shared representation. Fig. 5d shows that the task-specific electronic decoders used the same 16 optical channels to achieve 0.96 five-class accuracy. They achieved a 0.965 TPR at a 10% FPR for doublet detection and 0.949 SoftDice for morphology reconstruction. They achieved $R^2 = 0.926$ for morphology regression. The reconstructions retained the dominant cellular morphology, and the predicted descriptors followed the identity relation.

The matched comparison in Fig. 5e used identical 56×56 output sampling, 4×4 readout, and downstream decoder architectures. The comparison therefore probes the information content of the optical representation under matched electronic decoding capacity. Under these conditions, the optical-path-length-matched DNN benchmark achieved 0.86 accuracy and a 0.782 TPR. It also achieved 0.854 SoftDice for morphology reconstruction and $R^2 = 0.687$ for morphology regression. The OSNN exceeded the benchmark on all four evaluated metrics. Under the evaluated conditions, the OSNN representation supported higher performance with the same downstream decoder architectures.

## Discussion

The OSNN turns the optical volume into a trainable computing medium by co-distributing refractive-index neurons, optical transformations and interconnections throughout a monolithic three-dimensional structure. Its principal architectural advantage is that trainable degrees of freedom occupy the propagation volume rather than a sequence of separated surfaces. Replacing large axial gaps with densely distributed scattering sites yields a demonstrated neuron density of $1.0\times10^9$/mm$^3$. Because the computing medium is fabricated as one continuous structure, registration is intrinsic rather than dependent on assembly and alignment between planes. Across three experimental demonstrations, this same architecture supports class-selective energy routing, full-field compressed imaging and a shared physical representation for multitask single-cell analysis.

Under the tested conditions, expressive optical transformations required neither full connectivity nor strong modulation by individual neurons. Repeated local, angle-dependent interactions provide a physical inductive bias. Each update is weak and spatially concentrated, but depth progressively expands the effective coupling footprint and builds a global transformation. In classification and shared-representation tasks, the OSNN outperformed the corresponding diffractive benchmarks; in the classifier comparison, it also used a 78% shorter optical path. These results do not imply universal superiority of local connectivity, but show that dense weak interactions can use the available propagation volume efficiently. The calculated passive optical transit time is 0.82–0.93 ps. This intrinsic propagation time excludes input modulation, detector readout and electronic decoding. In the single-cell experiment, the measured output of the optical encoder co-designed for multiple tasks was pooled from 3,136 spatial samples into a 16-dimensional representation before task-specific decoding. Because the optical stage produces a shared representation rather than a task-specific decision, the same fabricated encoder can in principle support additional downstream tasks through newly trained electronic decoders, provided that the required information is retained in the optical latent space. Previous studies have embedded computation in wave dynamics and inverse-designed nanophotonic media as recurrent, matrix-multiplication, or inference processors[31–34]. Here, the OSNN experimentally distributes independently optimized refractive-index sites throughout a continuously occupied three-dimensional propagation volume, so the propagation space itself participates in the trainable computation and local transformation is physically co-located with optical interconnection. The current devices use binary 1-μm voxels and $\Delta n = 0.006$ for reproducible fabrication and efficient ordered weak-scattering modeling.

The present implementation also defines several directions for improvement. Grayscale index control, smaller voxels, a larger usable refractive-index contrast and lower optical loss could increase spatial bandwidth, volumetric density and transformation complexity. Incorporating a full multiple-scattering treatment into differentiable forward and gradient models could extend the architecture from the present weak-scattering regime to intermediate or strong scattering, while programmable materials could add reconfiguration. Supplementary Note 6 and Supplementary Fig. S6 extend the numerical model to mutually incoherent spectral–angular mode ensembles and show stable class-selective routing under a filled 1° source with a 20-nm bandwidth, motivating controlled-coherence experiments toward narrowband-LED, ambient-light and fluorescence-assisted sensing. At the system level, the current 60-Hz benchtop platform is dominated by SLM modulation, camera readout and electronic decoding rather than by passive propagation through the OSNN. For the nominal 60-Hz optical interface, the 56×56 transformation corresponds to $2.03\times10^{7}$–$5.90\times10^{7}$ real-operation equivalents per inference, giving a specification-based optical-interface estimate of 0.034–0.098 GOPS $W^{-1}$ under the stated component-energy assumptions (Supplementary Note 5, Section 5.5). Integrated input coupling, direct low-channel-count detection and co-designed electronic interfaces could substantially reduce this overhead and enable future measurements of end-to-end throughput, latency and wall-plug energy under application-scale conditions.

# Methods

## OSNN modelling and inverse design

The numerical implementation follows the weak-scattering, ordered-propagation formulation detailed in Supplementary Notes 1 and 2. The monolithic volume is discretized only numerically into $N$ virtual slices separated by $\Delta z$; these are sampling planes rather than fabricated layers. With transverse coordinate $\mathbf{r}_\perp = (x, y)$, complex field $E_s(\mathbf{r}_\perp)$ entering slice $s$, and slice scattering potential $V_s(\mathbf{r}_\perp)$, one propagation step combines nonparaxial background propagation with the first-order Born source field generated by the current slice. The slice-to-slice update is

$$E_{s+1} = \mathcal{P}_{\Delta z} E_s + \mathcal{G}_{\Delta z}[V_s E_s]. \tag{3}$$

Here, $\mathcal{P}_{\Delta z}$ denotes nonparaxial background propagation and $\mathcal{G}_{\Delta z}$ the first-order Born source propagator. In the spatial-frequency domain, $\mathcal{P}_{\Delta z}$ is evaluated with $H_{\Delta z}(\mathbf{f}_\perp) = \exp\{i\Delta z[(n_b k_0)^2 - (2\pi|\mathbf{f}_\perp|)^2]^{1/2}\}$, where $\mathbf{f}_\perp = (f_x, f_y)$, $n_b$ is the background refractive index and $k_0 = 2\pi/\lambda_0$. The outgoing complex square-root branch is selected so that propagating spatial frequencies acquire phase whereas evanescent components decay. Defining the pointwise multiplication operator $\mathcal{M}_{V_s}$ and the slice operator $\mathcal{A}_s = \mathcal{P}_{\Delta z} + \mathcal{G}_{\Delta z}\mathcal{M}_{V_s}$, the complete volumetric transformation is an axially ordered composition,

$$E_N = \mathcal{A}_{N-1}\mathcal{A}_{N-2} \cdots \mathcal{A}_1\mathcal{A}_0 E_0, \qquad \mathcal{A}_s \equiv \mathcal{P}_{\Delta z} + \mathcal{G}_{\Delta z}\mathcal{M}_{V_s}. \tag{4}$$

Optimization differentiates the real task loss $\mathcal{J}$ through this ordered complex-valued transformation. Using the Wirtinger adjoint field $\psi_s$, reverse propagation through $\mathcal{A}_s$ gives the sensitivity to every physical voxel. For voxel $j$ in slice $s$, whose transverse support is represented by $\chi_j$, the reverse recursion and refractive-index gradient are

$$\psi_s = \mathcal{A}_s^\dagger \psi_{s+1}, \qquad \frac{\partial \mathcal{J}}{\partial n_j} = 2\,\mathrm{Re}\big\langle \psi_{s+1}, \mathcal{G}_{\Delta z}\big[2k_0^2 n_j \chi_j E_s\big]\big\rangle. \tag{5}$$

The continuous material variables are optimized with Adam. Each physical voxel is assigned a real latent variable $u_j$; in the forward pass it is hard-projected to $n_{\mathrm{low}}$ or $n_{\mathrm{high}}$, whereas the backward pass uses the derivative of the smooth tanh relaxation through a straight-through estimator, as detailed in Supplementary Note 2. At every iteration, one common isotropic dimensional bias $\delta l_v \sim \mathcal{U}[-10,10]$ nm is applied to the voxel edge length and corresponding lattice pitch in all three dimensions, modeling a systematic global fabrication-scale error rather than independent voxel noise.

For the representative 56×56 devices, $N = 120$, $\Delta z = 1\,\mu\mathrm{m}$, $\lambda_0 = 532\,\mathrm{nm}$, $n_{\mathrm{low}} = 1.547$, $n_{\mathrm{high}} = 1.553$ and $\Delta n = 0.006$ on a 56×56×120 physical lattice. When used, an additional 60-μm homogeneous output-propagation region is evaluated with the same nonparaxial propagation formalism, giving a 180-μm total optical path. For MNIST classification, 28×28 inputs are replicated onto the 56×56 grid and ten fixed 4×4 detector regions provide the logits for a softmax cross-entropy objective; inference selects the region with the largest integrated intensity. The imager is optimized independently with an intensity-domain reconstruction loss on the 28×28 output, with FSIM reserved for evaluation. Thus, the classifier and imager share the same physical geometry and propagation model but learn different three-dimensional refractive-index distributions.

## Two-photon fabrication and structure

OSNN devices were fabricated by two-photon polymerization in IP-Dip2 negative-tone photoresist (Nanoscribe GmbH) on 2.5 cm×2.5 cm quartz substrates (refractive index approximately 1.46). Before writing, the substrates were rinsed with isopropyl alcohol (IPA) and dried with nitrogen. Three-dimensional printing was performed with a Photonic Professional GT2 system (Nanoscribe GmbH) in the dip-in laser lithography configuration. A 63× oil-immersion objective with a numerical aperture of 1.4 was immersed directly in the photoresist. The structures were written with a 780-nm femtosecond laser at a nominal power of 55 mW and a scanning speed of 10 mm $s^{-1}$. The optimized binary OSNN volume was converted to sliced writing patterns that specified the three-dimensional focal positions and therefore the locations of localized photopolymerization.

The functional optical measurements were performed after two-photon writing and before any solvent development. This sequence preserves the operating refractive-index contrast because the polymerized higher-index voxels remain embedded in the surrounding unexposed IP-Dip2 matrix, producing the weak refractive-index contrast between the unexposed resin and the exposure-modified resin used in the optical model. The classifier, imager and single-cell encoder were all measured in this undeveloped, resin-embedded state. Development was performed only after optical testing, or on separately prepared specimens used for structural characterization.

For post-test structural inspection, the samples were immersed sequentially in propylene glycol monomethyl ether acetate (PGMEA) for 5 min, IPA for 5 min and nonafluorobutyl methyl ether (NME) for 2 min. PGMEA removed the bulk unpolymerized photoresist, the IPA rinse displaced residual developer, and the final NME rinse removed remaining resist/developer residues before drying. To expose internal voxels, a developed structure was peeled from the quartz substrate with high-adhesion 3M VHB tape and inverted so that the bottom surface faced upward. Under an optical microscope, the tape-mounted structure was mechanically fractured with a microneedle to expose its central volume. Before scanning electron microscopy, the specimens were sputter coated with gold for 1 min at 40 mA. The internal morphology was then imaged with a JEOL JSM-7600F field-emission scanning electron microscope operated at 15 kV. These developed and fractured specimens provide the three-dimensional structural views shown in Figs. 3 and 4 and in the Supplementary Information, whereas all functional optical data were acquired before development.

**Optical measurements and readout**

The measurement system follows the optical path shown in Fig. 3a. A 532-nm laser beam is expanded and collimated, passed through a polarizer and directed by a beamsplitter onto a reflective phase spatial light modulator (FSLM-4K55-P02; 3840×2160 pixels; 3.2-μm pixel pitch). The modulated reflected beam is separated by the beamsplitter and passes through a second polarizer acting as an analyzer. In the crossed-polarizer amplitude-encoding configuration, the normalized transmitted intensity is proportional to the sine squared of half the SLM phase retardance; the beam is then relayed onto the OSNN input by a two-lens 4f system. The two lenses have focal lengths of 80 mm and 25 mm, respectively, corresponding to a transverse magnification of 0.3125. A Fourier-plane spatial filter is placed at the common Fourier plane between the two lenses. In this configuration, one 3.2-μm SLM pixel is mapped to 1.0 μm at the OSNN input plane, so a central 56×56-pixel SLM region maps directly onto the 56×56 $\mu m^2$ transverse aperture of the device. The SLM is positioned near the front focal plane of the first lens and the OSNN near the back focal plane of the second lens; the lens separation is set by the sum of the two focal lengths and fine-adjusted to maximize image registration and high-spatial-frequency contrast.

The OSNN output is collected with a 100× microscope objective and recorded on a CMOS camera. All measurements use the undeveloped devices described above. For MNIST classification, the intensity is integrated over the ten predefined detector regions and the predicted class corresponds to the region with the largest score. The classifier is evaluated on the complete 10,000-image MNIST test set. For compressed imaging, the same optical platform records the output intensity map produced by the separately optimized imager. Image fidelity is quantified by FSIM for 10,000 Fashion-MNIST test images and for 382 two-dimensional VesselMNIST3D-derived inputs evaluated with the same fixed OSNN without optical retraining. Spatial resolution is measured with the custom line chart in Fig. 4d. Two adjacent line responses are considered resolved when the intervening intensity minimum is no greater than 73.5% of the mean of the two neighboring peak intensities. The 1-μm group satisfies this criterion and defines the demonstrated imaging resolution. The optical configuration, detector definitions and evaluation metrics are detailed in Supplementary Note 5.

For the single-cell experiment, each measured 56×56 output intensity map is pooled over a fixed 4×4 grid to form a 16-dimensional physical latent vector. The same vector is supplied to all downstream tasks. Before trainable electronic decoding, it is expanded deterministically from 16 to 80 dimensions by a fixed, non-trainable feature transformation. This transformation provides the same nonlinear basis to each task but contains no learned parameters and therefore does not constitute a shared trainable electronic trunk. Four independently parameterized task-specific electronic decoders then perform organelle classification, morphology reconstruction, doublet detection and morphology regression. The matched diffractive benchmark uses the same detector sampling, the same fixed feature expansion and identical downstream decoder architectures, allowing task performance to be compared across optical encoders under matched readout and electronic decoding capacity. Full decoder architectures and parameter counts are given in Supplementary Table S2.

## Acknowledgements

H.D. acknowledges support from the National Natural Science Foundation of China (NSFC) under Grant Nos. W2611055 and 52425508. C.-W.Q. acknowledges support for this work from the Ministry of Education, Republic of Singapore (Grant Nos. A-8002152-00-00 and A-8004709-00-00), and the Competitive Research Program Award (NRF-CRP26-2021-0004 and NRF-CRP30-2023-0003) from the National Research Foundation, Prime Minister's Office, Singapore. This work was also supported by the National Key Research and Development Program of China (Grant No. 2023YFF0613600). We thank C. R. Chi, Y. Z. Xie and Z. Li for helpful suggestions and assistance with the simulation and experiment. We also thank Liang Shi for assistance with Blender-based rendering of the experimental setup.

## Author contributions

X.L., Q.S. and H.D. conceived the idea. C.-W.Q. and H.D. supervised the work. X.L. led the implementation and designed the models. X.L., Q.S., E.Y. and W.C. developed the mathematical derivations and trained the models. L.C., X.L., W.C., H.W. and Q.S. designed the experiments, fabricated the devices and performed the optical experiments. X.L., W.C., Q.S., C.-W.Q. and H.D. analyzed the data and wrote the manuscript. E.Y., L.C., Z.S., Y.H. and J.K.W.Y. contributed to writing and revising the manuscript. X.L., Q.S., W.C. and L.C. contributed equally to this work.

## Competing interests

The authors declare no competing interests.

## Additional information

Supplementary Information is available for this paper.

## Data availability

All raw data are available in the main text and the supplementary information.

## Code availability

The code will be made publicly available following acceptance.